\documentclass[twocolumn, usenatbib]{mnras}
\usepackage[utf8]{inputenc}
\usepackage{amssymb}
\usepackage{graphicx}
\usepackage{amsmath}
\usepackage{float}
\usepackage{subfigure}
\usepackage{hyperref}

\title[Exploring the consequences of chromatic data excision in 21-cm EoR power spectra]{Exploring the consequences of chromatic data excision in 21-cm Epoch of Reionization power spectrum observations}
\author[M. J. Wilensky et al.]{Michael J. Wilensky,$^{1, 2}$\thanks{m.wilensky@qmul.ac.uk} Bryna J. Hazelton,$^{2, 3}$ Miguel F. Morales,$^{2, 4}$
\\
$^{1}$Astronomy Unit Queen Mary University of London, Mile End Road, London, E1 4NS, United Kingdom\\
$^{2}$Department of Physics, University of Washington, Seattle, WA 98195, USA\\
$^{3}$eScience Institute, University of Washington, Seattle, WA 98195, USA \\
$^{4}$Dark Universe Science Center, University of Washington, Seattle, 98195, USA
}

\begin{document}

\maketitle

\begin{abstract}
    We explore how chromatic RFI flags affect 21-cm power spectrum measurements. We particularly study flags that are coarser than the analysis resolution. We find that such RFI flags produce excess power in the EoR window in much the same way as residual RFI. We use Fast Holographic Deconvolution (\textsc{FHD}) simulations to explain this as a result of chromatic disruptions in the interferometric sampling function of the array. We also use these simulations in conjunction with Error Propagated Power Spectrum with InterLeaved Observed Noise ($\varepsilon$\textsc{ppsilon}) to show that without modifying current flagging strategies or implementing extremely accurate and complete foreground subtraction, 21-cm EoR experiments will fail to make a significant detection. As a mitigation strategy, we find that circumventing the chromatic structure altogether by flagging the entire analysis band when RFI is detected is simple to implement and highly successful. This demands a detection strategy with a low false positive rate in order to prevent excessive data loss. 
\end{abstract}


\section{Introduction}

A promising probe to understand the Epoch of Reionization (EoR) is though the forbidden 21-cm emission from neutral Hydrogen. For reviews on the subject, see \citet{Furl2006}, \citet{Morales2010}, and \citet{Liu2019}. As can be gleaned from the above references, the EoR is very likely to have ended by a cosmological redshift of $z\approx6$, which places measurements of this probe firmly in the territory of low frequency radio telescopes. Current efforts to constrain models of reionization are focused on measuring the extremely faint 21-cm power spectrum signal.

Radio telescopes previously or currently involved in the search for the 21-cm EoR signal include the Murchison Widefield Array (MWA; \citet{Tingay2013}), The LOw Frequency ARray (LOFAR; \citet{LOFAR2013}), the Giant Metrewave Radio Telescope (GMRT; \citet{GMRT}), the Precision Array for Probing the Epoch of Reionization (PAPER; \citet{Parsons2010}), and the Hydrogen Epoch of Reionization Array (HERA; \citet{DeBoer2017}). These telescopes operate in unprotected radio bands, meaning they are subject to reception of anthropogenic radio signals, usually referred to in radio astronomy as radio frequency interference (RFI). These unwanted signals pose a significant threat to 21-cm EoR power spectrum measurements. This threat is quantified in \citet{Wilensky2020}, where it is estimated that a single RFI source of apparent flux density greater than 10 mJy in the final power spectrum integration can overwhelm a relatively optimistic EoR signal. \citet{Barry2019b} demonstrates that removal of faintly contaminated snapshots in power spectrum measurements has great potential to improve EoR power spectrum limits. Comparison of the results in \citet{Barry2019b} with those in \citet{Wilensky2020} suggest that improvements to RFI mitigation techniques are likely still necessary before residual RFI no longer poses a threat to EoR power spectrum measurements.

The first defense against RFI is to place a radio telescope in a relatively remote location. The difference that can be achieved is represented in Figure 10 of \citet{Offringa2015}, where RFI occupancy as a function of frequency as seen by AOFlagger \citep{Offringa2010, Offringa2012} is compared between MWA and LOFAR data. Most 21-cm EoR experiments additionally employ some form of post-correlation RFI excision pipelines, such as AOFlagger or SSINS \citep{Wilensky2019}. Such methods automatically identify or `flag' visibilities that are likely to be contaminated so that they are not used later in the analysis. 

\citet{Offringa2019a} explores some effects of excision methods in 21-cm EoR power spectrum analyses. This study is performed on LOFAR data and simulations, where RFI flagging is done at much finer spectral resolution than is used in the EoR analysis. The flagged visibilities are then downsampled to the analysis time and frequency resolution. Conventionally, flagged data are not included when downsampling, except when all contributing visibilities are flagged at high resolution. If there are some unflagged contributors, the resulting downsampled visibility is left unflagged at low resolution (referred to as partially flagged). Otherwise, the resulting visibility is flagged (referred to as fully flagged). Due to chromatic variations in flagging, conventional downsampling causes spectral fluctuations in the averaged visibilities that produces enough excess power in the EoR window to prevent detection of the expected EoR signal. This excess power can be effectively mitigated by improved downsampling schemes, weighting the downsampled visibilities identically regardless of how many samples contributed, subtraction of a low-resolution forward model, and Gaussian Process Regression (GPR: \citet{Kern2021, Mertens2018}). In this paper we primarily focus on fully flagged visibilities, which also produce spectral variations that result in excess EoR window power. Fully flagged and nearly fully flagged data is more common in the MWA EoR highband from 167-198 MHz than in the LOFAR band considered in \citet{Offringa2019a}. This is due to the presence of broad DTV interference in the MWA band as well as a lower correlator resolution. Some of the techniques from \citet{Offringa2019a} only address the excess power resulting from partially flagged visibilities (improved downsampling, reweighting), while others can address the excess power from fully flagged visibilities (forward modeling, GPR). We propose an alternate mitigation strategy that applies to fully flagged visibilities that is simple to implement and highly effective. We also investigate forward model subtraction.

In \citet{Hazelton2013}, the foreground wedge is understood in terms of the chromatic sampling function of the array, specifically by examining multi-baseline effects. The excess power in the window from fully flagged samples can be understood similarly. In \S\ref{sec:flagging_ps_methods}, we describe the theoretical concepts and methods used in this paper. In \S\ref{sec:demo} we use simulations to demonstrate and explain this effect, as well as provide a solution. In \S\ref{sec:data_flags}, we further demonstrate the effect using flags inherited from running \textsc{SSINS}  \citep{Wilensky2019} and \textsc{AOFlagger} \citep{Offringa2015} on MWA data. In \S\ref{sec:flag_ps_discussion}, we discuss mitigation strategies and consequences for overall experimental sensitivity.

\section{Theoretical Principles and Methods}
\label{sec:flagging_ps_methods}

The essential theoretical principle that explains excess power from fully flagged visibilities is that they cause sharp spectral variations in the sampling function of the array. This causes power from spectrally smooth foregrounds, which would otherwise be contained in the wedge, to leak up into the EoR window. The shape of the leaked power is directly related to the shape of the spectral variations via a Fourier transform. The overall amplitude is proportional to the total foreground power. Since the foregrounds are tremendously bright compared to the expected EoR signal, even small variations caused by a minute fraction of fully flagged samples cause enough power in the EoR window to prevent detection. 

Due to the rotation of the Earth, the $uv$-modes sampled by a terrestrially bound interferometer depend not only on the array layout and the observing frequencies, but also on the sidereal times present in the data. For instance, if the phase center is held constant, a projected baseline center for a given pair of antennas and frequency tracks an ellipse in the $uv$-plane. This effect is typically utilized to sample more modes than would otherwise be included by a given antenna pair in a technique called rotation synthesis. If LST sampling is even, the average location of a given baseline will be halfway along the arc of the ellipse that is swept out during the range of LSTs. If LST sampling is uneven, then the average location of the baseline shifts away from this center. Extending this logic, if the set of sampled LSTs for one frequency is different than that of another frequency, the average locations of the baseline centers will possess spectral variations that are potentially very sharp. The disappearance of baselines for some frequencies will also imprint these variations even if the LST-averaged location of the baseline is frequency-independent. This is the primary effect on the sampling function that causes the excess power in the EoR window. Note that these spectral variations are present regardless of whether the data are gridded before power spectrum estimation.  This means that any analysis that coherently time-averages measurements can have excess power as a result of flagging, and thus the results in this work apply equally well to all types of power spectrum analyses \citep{Morales2019}.

The spectral shape of the sampling function variation is directly influenced by the shape of the flags. This shape subsequently determines the form of the power spectrum contamination. There are five basic flagging shapes in the MWA EoR highband, resulting from different sources: 
\begin{enumerate}
    \item Narrowband (single-frequency transmitters)
    \item Broad, band-limited (DTV)
    \item Frequency Comb (routine coarse band edge flagging)
    \item Uniform Random (false positives from the flagging algorithm)
    \item Broad, not band-limited (various, e.g. extremely bright ORBCOMM spillover)
\end{enumerate}
The spectral variation introduced by any given flagging shape is identical to the spectral variations introduced by leaving an unflagged RFI source that occupies the same frequencies as those that are flagged, but with a strength given by the foreground brightness and flagging occupancy rather than the strength of the RFI. Therefore, we can understand the form of the excess window power in terms of the analysis presented in \citet{Wilensky2020}. Narrowband flags produce constant power in the line-of-sight modes, while DTV flags produce lobed power-law contamination. A frequency comb of flags produces a line-of-sight comb in the power spectrum, while uniformly random flags produce uniformly random contamination. Finally, totally achromatic flags, equivalent to flagging entire integrations with any RFI, produces no excess power in the window.

Flags tend to be irregular in terms of sidereal times since most RFI is not locked to a sidereal schedule. Consequently, various modes experience chromatic disruptions in sampling, even if the physical baselines that are flagged are consistent. To demonstrate the resulting excess power, we simulate MWA visibilities with various flags applied in a number of circumstances and calculate the resulting power spectra using the \textsc{FHD}/$\varepsilon$\textsc{ppsilon} pipeline \citep{Barry2019a, Sullivan2012}. While this is a gridded power spectrum pipeline, we expect that the effect will still appear in delay spectrum pipelines as well. Unless a given baseline is flagged perfectly consistently as a function of frequency and LST, that baseline will sample its rotation ellipse irregularly as a function of frequency, and this will throw power into high delay modes. 

%To explain and explore this concept, we use a suite of simulations with simple flagging patterns on two minutes of visibilities. Starting with a single source that has been flagged for DTV, we show that chromatic flags result in excess power. In order to get a more realistic estimate of the strength of the effect, we then simulate two minutes of visibilities using a subset of the \textsc{GLEAM} catalog, and explore more flagging patterns. Finally, to understand how the effect will appear in measured power spectra, we simulate a two-hour observing period with flags inherited from actual MWA data. 

\section{Conceptual Demonstration}
\label{sec:demo}

In this section, we simulate two minutes of MWA visibilities from two different source catalogs and various flags using \textsc{FHD}. First, we examine the gridded, flagged visibilities of a single source displaced from phase center, which demonstrate the sharp spectral fluctuations in the reconstructed $uv$-plane that create the excess power. We then work through several simulations of a \textsc{GLEAM} \citep{Hurley-Walker2017} subset using the flagging shapes listed in \S\ref{sec:flagging_ps_methods} to give a lower bound on the strength of this effect in measured power spectra. The simulations are centered on a right ascension of 0 and declination -27 degrees. For all power spectra shown in this work, we use the entire 30.72 MHz instantaneous bandwidth of the simulated visibilities. In practice, only a fraction of this bandwidth is used so as to ensure that only nearby epochs are probed, i.e. so that the measurement does not span too wide a redshift. We find the power spectrum features arising from chromatic flags are more clearly resolved using a wider bandwidth, and so we use the full band for the sake of demonstration.

\subsection{A Single Flagged Source}
\label{sec:one_source}

% This will demonstrate that chromatic flagging causes excess power at high $k_\parallel$ modes as a result of sharp spectral variations in the $uv$-sampling function. 

\begin{figure}
    \centering
        \includegraphics[width=0.9\linewidth]{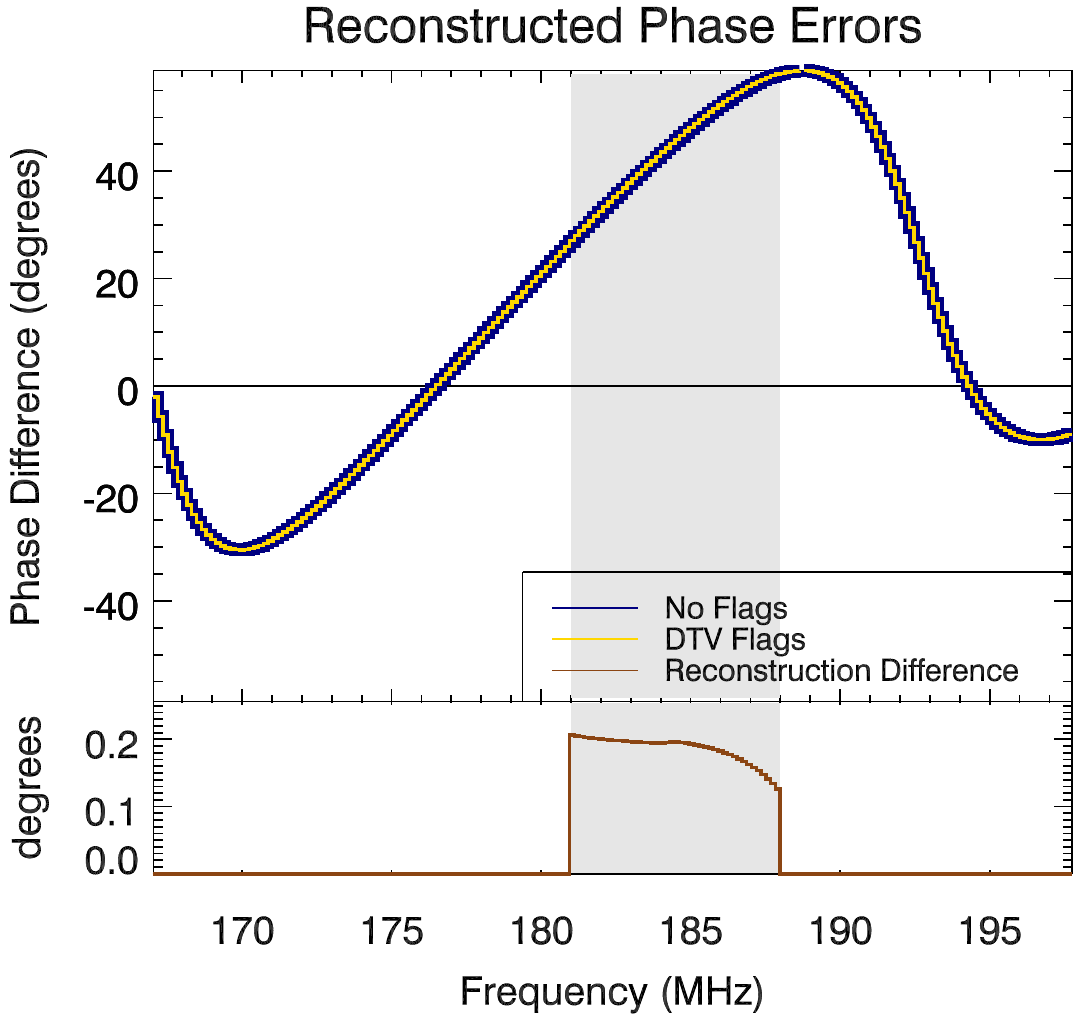}
    
    \caption{Top: Shows difference between analytic phase of a source at a given $uv$-point and the reconstructed phase using \textsc{FHD} with no flags (dark blue), and with DTV flags (gold). The smooth spectral variation in the phase reconstruction error is what gives rise to the foreground wedge. Bottom: Phase difference between the reconstructed phase with minimal flagging and with DTV flags. There is only disagreement where flags between the two reconstructions disagree, and the disagreement appears discontinuously at the boundaries of the flagged region. This sharp spectral structure gives rise to power outside the wedge, in the EoR window, cf. Figure \ref{fig:multibaseline_ps}.}
    \label{fig:one_source}
\end{figure}

We begin by examining the $uv$-plane resulting from a single point-source displaced from phase center under different flagging conditions. We consider three quantities:
\begin{enumerate}
    \item The analytic $uv$-plane.
    \item The reconstructed $uv$-plane using \textsc{FHD}, with no flags.
    \item The reconstructed $uv$-plane using \textsc{FHD} as above, but with flags from 181-188 MHz that simulate digital television excision. The extra flags are on all baselines for half the integrations in the simulation.
\end{enumerate}
In Figure \ref{fig:one_source}, we examine the difference between the analytic phase and the two reconstructed phases enumerated above at a single $uv$-point. The topmost panel of Figure \ref{fig:one_source} shows the difference between the analytic phase and the reconstructed phase with no flags applied, as well as the difference between the analytic phase and the reconstructed phase using DTV flags. Over the range of frequencies in the band, both residuals appear smoothly varying and nearly identical. The large-scale, smooth, but nonzero variation in the reconstructed phase error is what gives rise to the foreground wedge, via the mechanism explained in \cite{Hazelton2013}. The bottom panel shows the difference between the two reconstructed phases. There is only disagreement for frequencies with flags, and the fractional discrepancy in the phase is $\sim10^{-3}$. This small discontinuity accounts for power outside the wedge in the EoR window. 

\begin{figure}
    \centering
    \includegraphics[width=0.8\linewidth]{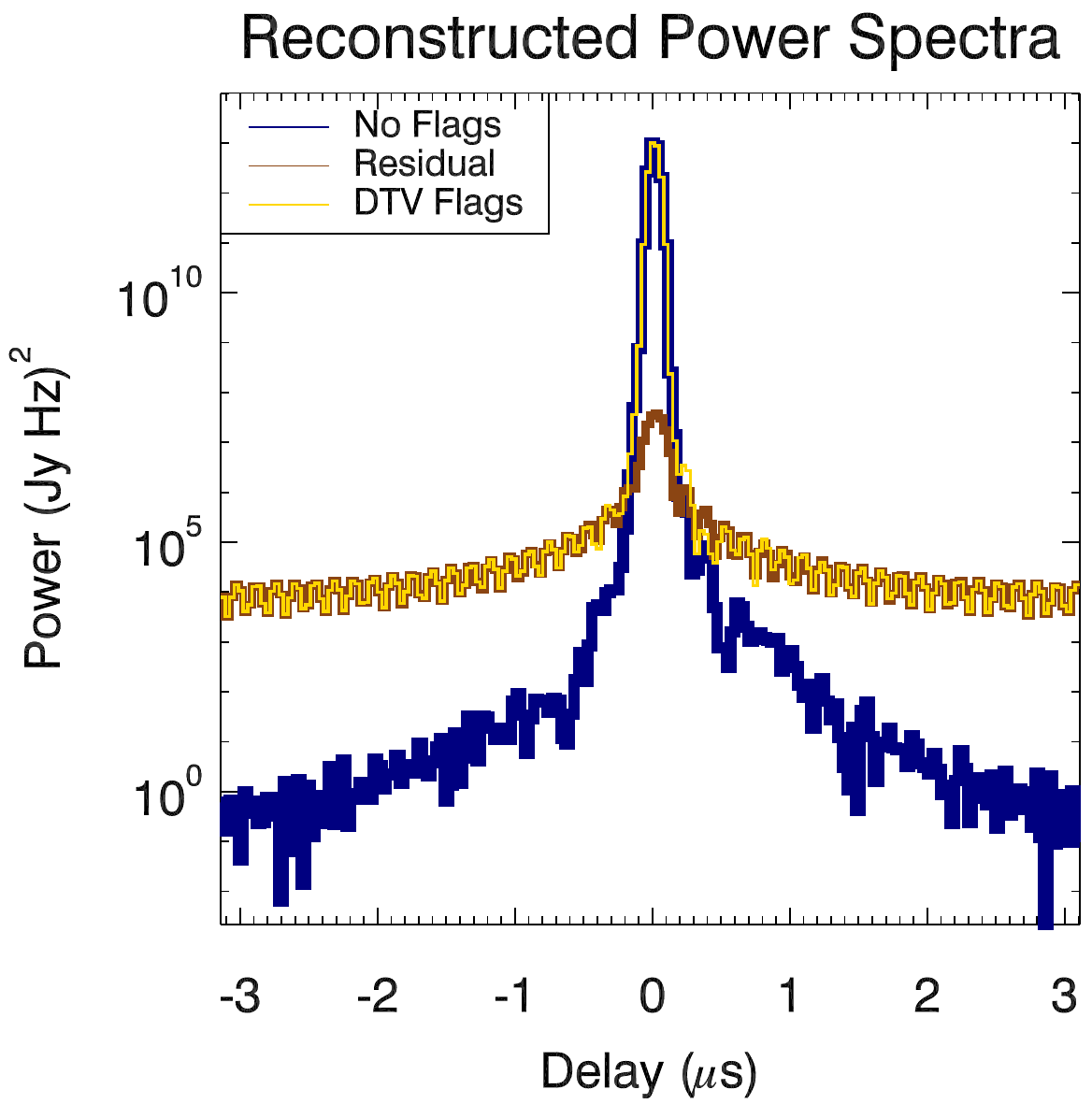}
    \caption{Power spectra of the two reconstructed $uvf$-cubes and their residual, evaluated at a single $uv$-point. The power spectrum with no flags falls off at a similar rate as a Blackman-Harris window function's Fourier transform, as would be expected for a smooth-spectrum source. Adding flags gives a similar amount of power in the zero-delay mode, but with much slower falloff. The power spectrum of the residual is low in the zero-delay mode since the discontinuity is small, however it has long tails that match the tails of the flagged reconstruction.}
    \label{fig:multibaseline_ps}
\end{figure}

To show how this affects the power spectrum, we evaluate the modulus square of a windowed spectral Fourier transform of the reconstructed $uvf$-cubes at the $uv$-point considered in Figure \ref{fig:one_source}, using a Blackman-Harris spectral tapering function. We then difference the $uvf$-cubes (equivalent to differencing the visibilities before gridding) to form a `residual' cube, and compute its power spectrum in the same way. We show the results in Figure \ref{fig:multibaseline_ps}. We find that the flagged reconstruction has long tails out to high-order delay line-of-sight modes, while the unflagged reconstruction has favorable falloff. The tails of the residual power spectrum match the tails of the flagged power spectrum. The rate of falloff for the flagged data is exactly what would be expected for a sharp discontinuity, similar to the RFI power spectra in \citet{Wilensky2020}.

The size of the discontinuity is consistent with the relative baseline center offset between flagged and unflagged data. The discrepancy in baseline position can be understood by examining the trajectory of a baseline center as a function of time. The arcspeed of a baseline along its elliptical track is fastest when it its projected displacement vector is aligned with $u=0$. For intervals of fixed length and a given baseline, centering observations on this LST produces the greatest discrepancies between flagged data and unflagged data. For such baselines and times, the arc length, $\Delta s$, swept out during a short time interval is approximately
\begin{equation}
    \Delta s \approx X\Delta\tau,
\end{equation}
to first order in $\Delta\tau$, where $X$ is the East-West length of the baseline in question when projected from zenith at the given frequency, and $\Delta\tau$ is the length of the LST interval, measured in radians. For the unflagged data in the simulation, the average baseline locations are at the center of their respective arcs. For the flagged data, the average baseline centers are located away from the center of this arc. Since the first half of the two-minute interval is flagged, the average location of each baseline is approximately three-quarters of the way along the arc. This discrepancy is 
\begin{equation}
    \Delta s_{\text{max}} \approx \frac{\pi}{1440}X \sim (2 \cdot 10^{-3})X
    \label{eq:arcdist}
\end{equation}
A given baseline's arcspeed will be slowest when it projects along $v=0$. The factor by which it is slower than its maximum speed is given by $\sin\delta_0$, where $\delta_0$ is the declination of phase center. For the MWA EoR0 field, centered on a right ascension of 0 and declination of -27 degrees, this is about a factor of 2. Figure \ref{fig:one_source} shows that this discrepancy produces a phase difference between the unflagged and flagged, gridded visibilities on the order of 0.2 degrees that is spectrally discontinuous at the borders of the flagged regions. We note that the relative size of this effect matches the discrepancy in power between the unflagged and residual power spectra in the zero-delay mode; the discrepancy in zero-delay power is fractionally $3.39 \cdot 10^{-6}$, which is slightly smaller\footnote{This is to be expected, since equation \ref{eq:arcdist} represents the maximum amount and the baselines in question do not project along $u=0$ in the simulation.} than the square of the size of the fractional baseline displacement for that particular $uv$-point based on equation \ref{eq:arcdist}.

A discontinuity in the frequency domain results in power law contamination in the Fourier domain that is proportional to the size of the discontinuity. In \citet{Wilensky2020} this was demonstrated by RFI sources introducing discontinuities. Here, it is the flags that produce the discontinuity, but the overall effect is the same. In practice, the sky signal is dominated by the foregrounds, and the size of the discontinuity resulting from the aberrant sampling function is proportional to their brightness. Since the foreground signal is orders of magnitude brighter than the EoR signal, the resulting contamination in the EoR window is significant. In the next section, we simulate visibilities and power spectra for \textsc{GLEAM} sources to get a rough estimate of the strength of this effect in a measured power spectrum. Since no source catalog is complete, and we do not model diffuse emission, the strength of this effect in simulated power spectra serves as a lower bound for the strength in measured power spectra.

\subsection{Flagged GLEAM}
\label{sec:flagged_GLEAM}

\begin{figure*}
    \centering
    \begin{subfigure}[]{
    \includegraphics[scale=0.8, page=2]{flagging_PS_figs/flagging_PS_2_minutes_basic.pdf}
    }
    \end{subfigure}
    \centering
    \begin{subfigure}[]{
    \includegraphics[scale=0.8, page=1]{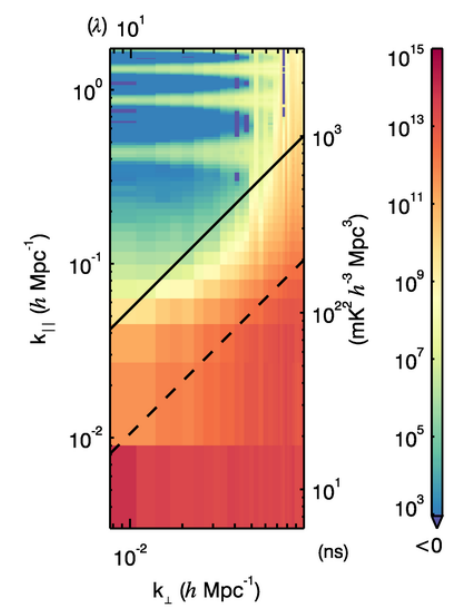}
    }
    \end{subfigure}
    
    \centering
    \begin{subfigure}[]{
    \includegraphics[scale=0.8, page=4]{flagging_PS_figs/flagging_PS_2_minutes_basic.pdf}
    }
    \end{subfigure}
    \centering
    \begin{subfigure}[]{
    \includegraphics[scale=0.8, page=3]{flagging_PS_figs/flagging_PS_2_minutes_basic.pdf}
    }
    \end{subfigure}
    \caption{Simulated power spectra using GLEAM and different sets of flags: (a) no flags, (b) coarse band edge flags only, (c) random flags in addition to coarse band edge flags, and (d) 7 MHz corresponding to DTV channel 7 flagged in addition to coarse band edge flags.}
    \label{fig:flag_PS_sim_GLEAM}
\end{figure*}

In this subsection, we repeatedly simulate two minutes of visibilities from sources in the \textsc{GLEAM} catalog that lie in the MWA field of view centered on a right ascension of 0 and declination of -27 degrees using \textsc{FHD}. We include a beam that uses an average embedded element and mutual coupling model \citep{Sutinjo2015}. We simulate a single beam, and therefore ignore any potential effects from beam chromaticity. In each new simulation we use the same sources and observing parameters, but change the flagging pattern. The simulated data are flagged and gridded at 80 kHz. Then HEALPix cubes are formed and averaged to 160 kHz for passage to $\varepsilon$\textsc{ppsilon}. This typically results in a mixture of fully-flagged and partially-flagged visibilities. Figure \ref{fig:flag_PS_sim_GLEAM} shows the simulated power spectra of the \textsc{GLEAM} sources with different flags applied. There are four chosen flagging patterns: no flags, coarse band edge flags only, random flags overtop coarse band edge flags, and DTV flags overtop coarse band edge flags. Each flagging pattern produces a characteristic shape in the EoR window. The purpose of the latter two patterns is to simulate what is found in MWA data. The case without flags sets the ground truth for the GLEAM power spectrum with our current analysis techniques. The window is free of foreground contamination, with a floor given by the Fourier transform of the various tapering functions applied to the data \citep{Barry2019b}.

Coarse band edge flags are periodic in frequency. We flag 80 kHz on either side of each coarse channel edge. This means that the first and last fine 80-kHz channels of each coarse channel are flagged. There are an even number of fine channels per coarse channel. When we average to 160 kHz, the resulting edge channels include data from one flagged channel and one unflagged channel, assuming no additional flags have been added. In other words, the resulting 160 kHz coarse band edges of the \textsc{HEALPix} cubes are always at least half-flagged. This produces a harmonic structure in the Fourier domain along the line-of-sight modes. All power spectra with coarse band edge flags will have this harmonic structure present in them, and it is several orders of magnitude brighter than the expected EoR signal strength. Each harmonic has a non-negligible width in the line-of-sight modes, and affects all baselines, making a wide range of modes in the EoR window totally inaccessible.

The bottom-left panel of Figure \ref{fig:flag_PS_sim_GLEAM} shows the results of applying random flags to the GLEAM visibilities at a rate of approximately 0.2 per cent, in addition to coarse band edge flags. This simulates the effect of a small false positive ratio when flagging at the data resolution. \textsc{AOFlagger} is applied to MWA data at the finest resolution possible in order to have the greatest detection accuracy. The actual false positive rate at the finest resolution is higher than this rate. This rate reflects the number of fully flagged samples that typically remain after averaging from 40 kHz to 80 kHz, and ignores effects from partially flagged samples. These flags are distributed uniformly across all times, baselines, and non-edge frequencies. We see that for a two-minute simulation, this causes excess power with a noisy shape that is many orders of magnitude above the expected EoR signal. 

The bottom-right panel of Figure \ref{fig:flag_PS_sim_GLEAM} shows the results of flagging all baselines for the first half of the observation across all baselines for the frequencies 181-188 MHz. This pattern might be expected from running \textsc{SSINS} on an observation with DTV contamination resulting form aircraft reflection, as in \citet{Wilensky2019}. This produces a lobed structure as a function of $k_\parallel$, and in fact the size of the lobes corresponds to the inverse-width of the flagged region, as would be expected from basic Fourier transform reasoning and the calculations shown in \citet{Wilensky2020}.\footnote{The lobes in this power spectrum appear slightly different than those in the simulated DTV power spectrum in \citet{Wilensky2020}. This is a result of having chosen a different analysis bandwidth, which affects the width of the $k_\parallel$ bins for which the spectrum is calculated.} This contamination goes approximately as $k_\parallel^{-2}$. Extrapolating this back to the lowest order $k_\parallel$ mode, we see that this is roughly 5-6 orders of magnitude beneath the GLEAM power, which corresponds to the excess power amplitude being quadratic in the sampling function discontinuity as in the single-source case. The contamination is also several orders of magnitude above the expected EoR signal strength in the window. The spacing of the lobes makes measurements in the window impossible without mitigation of this effect. 

\subsection{LST Replacement}
\label{sec:LST_replace}

We can associate a given sampling configuration with a local sidereal time and frequency combination. If missing LST/frequency combinations can be appropriately replaced by data from another night, then the sampling function can be restored to its original smoothness. Unfortunately, we find that naively averaging flagged data with corresponding unflagged data generically results in only a minor reduction of excess window power if special care is not taken. 

\begin{figure}
    \centering
    \begin{subfigure}{
    \includegraphics[page=1, scale=0.5]{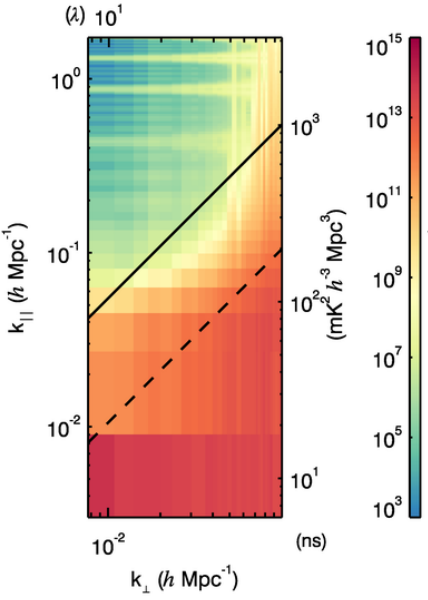}
    }
    \end{subfigure}
    \begin{subfigure}{
    \includegraphics[scale=0.2658]{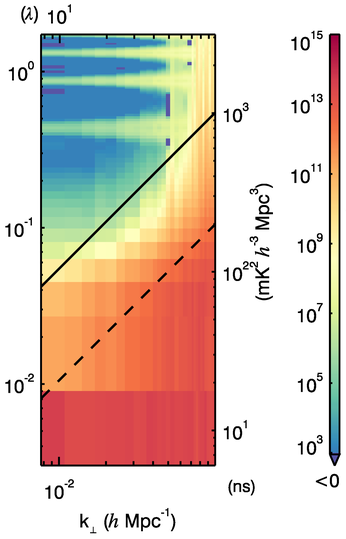}
    }
    \end{subfigure}

    \caption{(a) Simulated power spectrum formed by flagging the first half of a two-minute interval with 7 MHz DTV flags and averaging it against the exact same observation but without any flagging. EoR window contamination levels are only mildly reduced and still too high. (b) Simulated power spectrum where the first half of the two minute interval is flagged across all frequencies, rather than just those corresponding to the DTV allocation. Coarse band edge flags are still present in the second half of the observation. No excess power in the window other than what arises from coarse band edge flags is added by this flagging strategy.}
    
    \label{fig:two_night_sim}
\end{figure}

To investigate averaging flagged nights with unflagged ones, we simulate two copies of a two-minute MWA observation: one with DTV flags as in Figure \ref{fig:flag_PS_sim_GLEAM}(d), and another with no flags. We then coherently average the resulting HEALPix cubes to observe the effect of exact LST/frequency replacement. Relative to other LST/frequency combinations, the flagged LST/frequencies are still undersampled after the coherent average. While this reduces the strength of the discontinuity, it does not eliminate it. In Figure \ref{fig:two_night_sim}(a), we show the resulting power spectrum of the coherently averaged cubes. The excess power is only slightly diminished. 

Curiously, if we create a copy of the two-minute interval with exactly complementary flags, i.e. with the second half flagged for DTV, and average it with the first, this produces a weighted average whose $uv$-plane is smooth as a function of frequency. The resulting power spectra have no excess window power. While we recognize that this method successfully removed the excess power resulting from fully flagged visibilities, this would be a costly mitigation strategy in terms of sensitivity. Since exactly complementary flags are required, this reduces the overall data volume of an analysis by at least a factor of two.

\subsection{Achromatic Flags}
\label{sec:achromatic_flags}

We observe that if we flag all frequencies for any integration that is contaminated, then the window contamination is also removed. Since this is equivalent to simply not including any contaminated integrations, no additional chromaticity is injected into the $uv$-sampling function, resulting in no additional power other than what exists from the routine coarse band edge flagging. The ensuing power spectrum is shown in Figure \ref{fig:two_night_sim}(b). This potential mitigation strategy has an associated sensitivity cost that depends crucially on the flagging strategy. We propose this as the primary strategy for mitigating excess power from RFI flagging so long as the false positive rate of the flagger is sufficiently low. In the next section, we assess the impact of this mitigation strategy using simulations with flags inherited from measured data. 

\section{Flags Inherited from Data: Assessment and Mitigation}
\label{sec:data_flags}

In this section, we examine the strength of the excess flagging power using one night of flags inherited from MWA data, as well as the impact of the mitigation strategy alluded to in the preceeding section. We also check to see if flagging power significantly reduces over the course of a night by analyzing model power spectra from in-situ simulations of GLEAM using the MWA as the dummy instrument, with flags calculated by \textsc{SSINS} and \textsc{AOFlagger} on just over two hours of measured MWA visibilities. The data are flagged at high resolution (2-seconds, 40 kHz) and theresulting data is downsampled to 80kHz. The overall \textsc{SSINS} occupancy at high resolution is 4.4 per cent, ignoring coarse band edges and the temporal boundaries of observations. Under the same conditions, the \textsc{AOFlagger} flags about 2 per cent of the data at high resolution.

\subsection{Integrating Over a Night}

\begin{figure}
    \centering
    \includegraphics[scale=0.2]{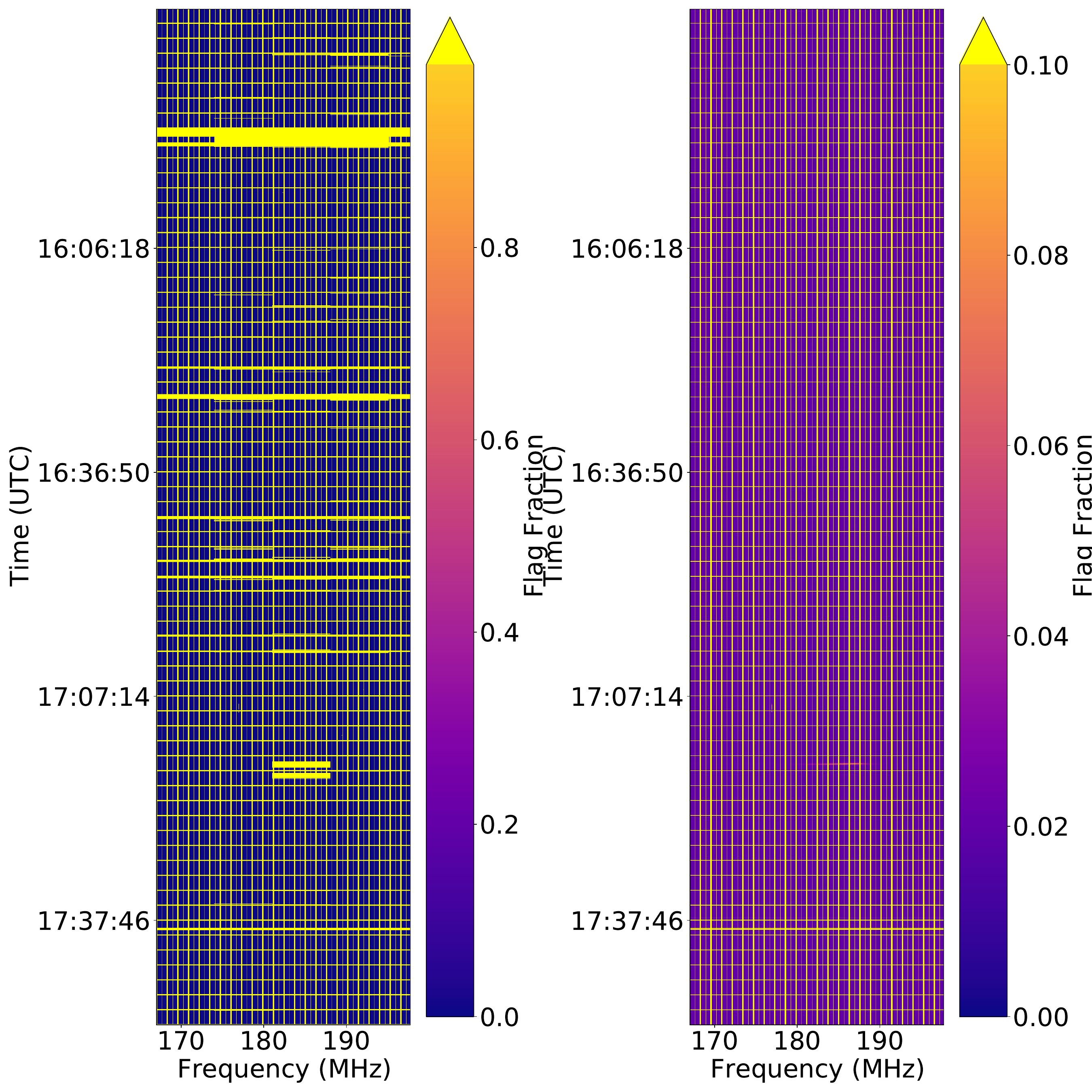}
    \caption{The fraction of baselines flagged using \textsc{SSINS} (left) and \textsc{AOFlagger} (right). Yellow-colored samples are flagged across all baselines. The vertical axis tick marks represent pointing boundaries. Since \textsc{SSINS} averages over baselines, either all baselines are flagged at a given time and frequency or none of them are. \textsc{SSINS} was set to specifically search for DTV, hence the somewhat regular 7 MHz blocks. \textsc{AOFlagger} tends to flag uniformly in the band, in part due to a small but noticeable false positive ratio, except some short DTV and narrowband events around 17:16 and 17:09 UTC, respectively.}
    \label{fig:gold_flagging_pattern}
\end{figure}

\begin{figure}
    \centering
    \includegraphics[scale=0.2]{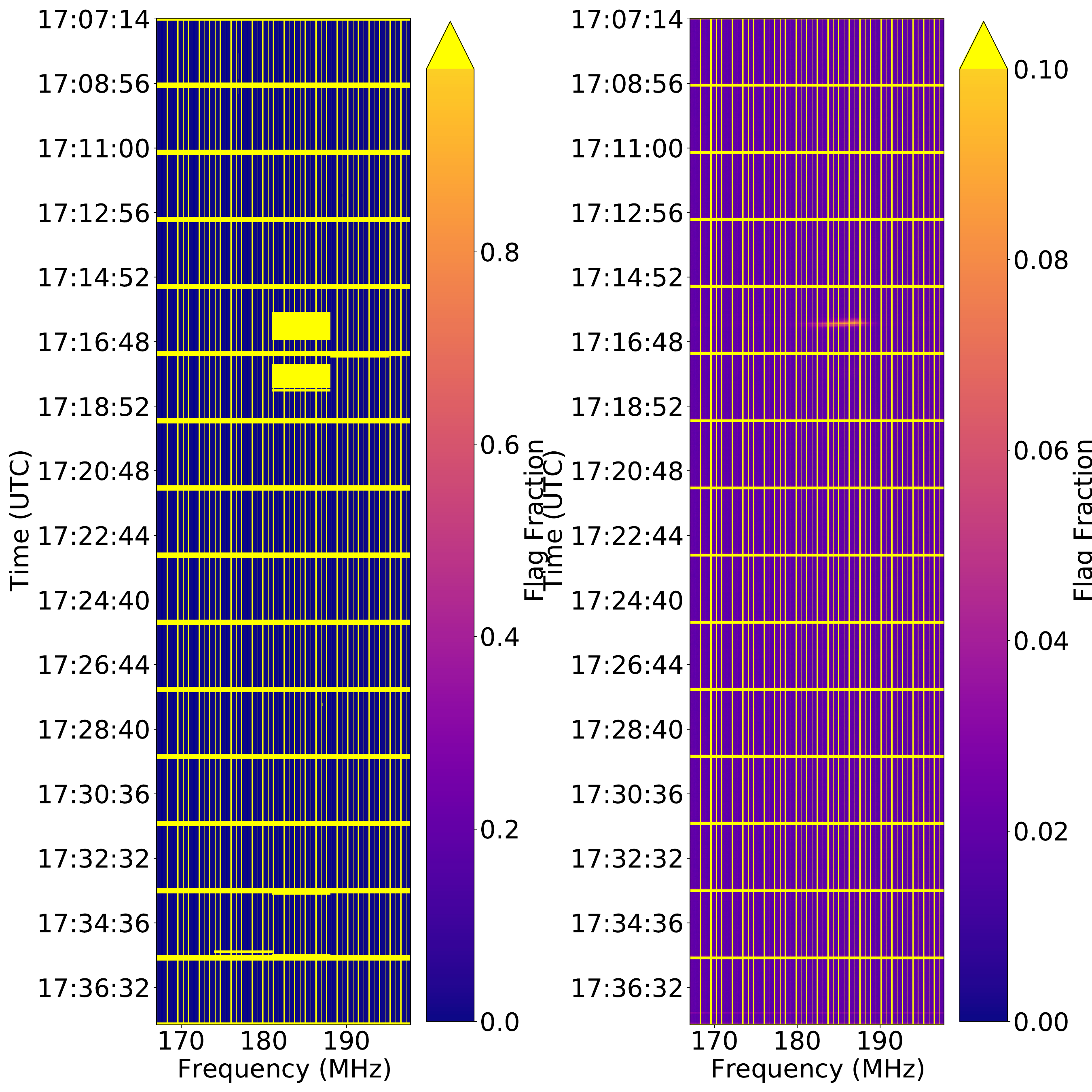}
    \caption{Closer view of the \textsc{SSINS} (left) and \textsc{AOFlagger} (right) flags from the fourth pointing in the night. Both \textsc{SSINS} and \textsc{AOFlagger} observe a narrowband event shortly before 17:09 UTC at about 177 MHz. There is also a DTV event around 17:16 UTC, due to an airplane reflection that both flaggers catch, though \textsc{AOFlagger} identifies a fewer contaminated times, and does not identify the event in the following observation. Very close to the end of the pointing, \textsc{AOFlagger} identifies a broadband event on a few baselines that \textsc{SSINS} does not identify. It is possible that this event is sufficiently diluted by the incoherent average in \textsc{SSINS} to become undetectable. While broadband flagging actually does not introduce additional chromaticity into the measurement since all frequencies are flagged equally, it is important to acknowledge possible false negatives in the \textsc{SSINS} pipeline.}
    \label{fig:plus_one}
\end{figure}

We show the baseline-averaged flagging patterns for the two-hour data set in Figure \ref{fig:gold_flagging_pattern}. The periodic flagging in time and frequency are not strictly RFI-related. The MWA coarse channel edges and centers are inhabited by bandpass irregularities arising from the two-stage polyphase filter bank in the digital signal pipeline that need to be excluded from the analysis. The time boundaries of observations also tend to have irregularities due to misalignment of data coming from the correlator and occasional oddities when pointings are adjusted, among other effects. The regular 7 MHz blocks in the \textsc{SSINS} flags reflect the results from the frequency-matched flagging algorithm, which specifically searched for DTV contamination. The flags from 
\textsc{AOFlagger} are mostly uniform except for noticeable DTV and narrowband events later in the night, corroborated by the \textsc{SSINS} flags. We show a closer view of these events in Figure \ref{fig:plus_one}. Since \textsc{AOFlagger} flags on a per-baseline basis and produces a small number of false positives, each baseline actually has a slight, random chromaticity relative to other baselines. Each of these types of chromatic structures, namely coarse band edge flagging, DTV, and uniform random, imprint visible structures in the EoR power spectrum according to their Fourier transform.

\begin{figure*}
    \centering
    \begin{subfigure}[]{\includegraphics[width=0.3\linewidth, page=3]{flagging_PS_figs/SSINS_2014_GS_pointing_PS_cropped_titles.pdf}
    }
    \end{subfigure}
    \begin{subfigure}[]{\includegraphics[width=0.3\linewidth, page=2]{flagging_PS_figs/SSINS_2014_GS_pointing_PS_cropped_titles.pdf}
    }
    \end{subfigure}
    \begin{subfigure}[]{\includegraphics[width=0.3\linewidth, page=6]{flagging_PS_figs/SSINS_2014_GS_pointing_PS_cropped_titles.pdf}
    }
    \end{subfigure}
    \begin{subfigure}[]{\includegraphics[width=0.3\linewidth, page=4]{flagging_PS_figs/SSINS_2014_GS_pointing_PS_cropped_titles.pdf}
    }
    \end{subfigure}
    \begin{subfigure}[]{\includegraphics[width=0.3\linewidth, page=5]{flagging_PS_figs/SSINS_2014_GS_pointing_PS_cropped_titles.pdf}
    }
    \end{subfigure}
    \begin{subfigure}[]{\includegraphics[width=0.3\linewidth, page=1]{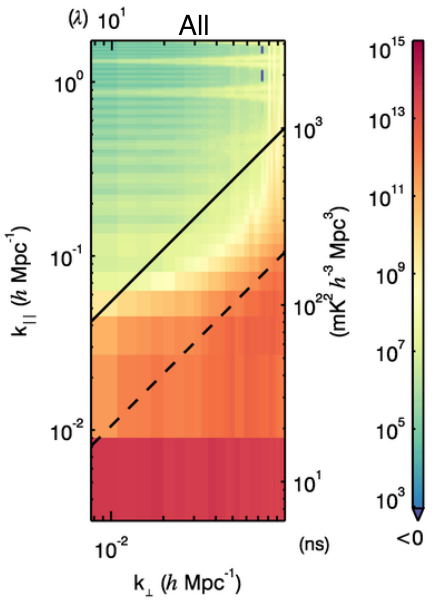}
    }
    \end{subfigure}
    \caption{Simulated per-pointing power spectra using \textsc{SSINS} flags inherited from data. The five pointings are shown in chronological order left-to-right, top-to-bottom. There are designated with integers ranging from -2 to 2 above each spectrum. They progress from East to West as the array tracks a field, 0 being zenith. The sixth power spectrum is the result from integrating the entire night. The fourth pointing is substantially worse than the others due to a bright DTV event that produced many DTV flags. The total integrated power spectrum contains contamination in the window that overwhelms the expected EoR signal by several orders of magnitude.}
    \label{fig:model_flag_PS_data_SSINS}
\end{figure*}

\begin{figure*}
    \centering
    \begin{subfigure}[]{\includegraphics[width=0.3\linewidth, page=3]{flagging_PS_figs/COTTER_2014_GS_cropped_titles.pdf}
    }
    \end{subfigure}
    \begin{subfigure}[]{\includegraphics[width=0.3\linewidth, page=2]{flagging_PS_figs/COTTER_2014_GS_cropped_titles.pdf}
    }
    \end{subfigure}
    \begin{subfigure}[]{\includegraphics[width=0.3\linewidth, page=6]{flagging_PS_figs/COTTER_2014_GS_cropped_titles.pdf}
    }
    \end{subfigure}
    \begin{subfigure}[]{\includegraphics[width=0.3\linewidth, page=4]{flagging_PS_figs/COTTER_2014_GS_cropped_titles.pdf}
    }
    \end{subfigure}
    \begin{subfigure}[]{\includegraphics[width=0.3\linewidth, page=5]{flagging_PS_figs/COTTER_2014_GS_cropped_titles.pdf}
    }
    \end{subfigure}
    \begin{subfigure}[]{\includegraphics[width=0.3\linewidth, page=1]{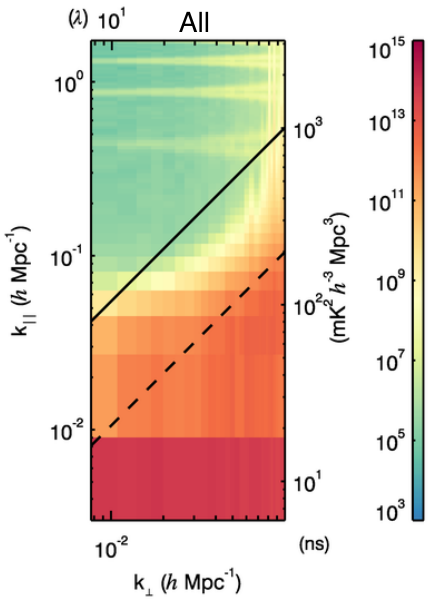}
    }
    \end{subfigure}

    \caption{Simulated per-pointing power spectra using \textsc{AOFlagger} flags inherited from data. The five pointings are shown in chronological order left-to-right, top-to-bottom. See Figure \ref{fig:model_flag_PS_data_SSINS} caption for pointing designations. The sixth power spectrum is the result from integrating the entire night.  While \textsc{AOFlagger} did catch the DTV event in the fourth pointing on some baselines, the effect is not as discernible in these spectra as it was in the \textsc{SSINS}-flagged spectra. The EoR window contamination is still too high for a significant EoR detection.}
    \label{fig:model_flag_PS_data_AOFlagger}
\end{figure*}

Individual dipole delays within an MWA tile can be adjusted so that the tile can point its beam. For EoR experiments with the MWA, this functionality is used to track a particular observing field as it rotates overhead. In Figure \ref{fig:gold_flagging_pattern}, we demarcate these transitions with the time axis tick marks. They occur about every 30 minutes. We show model power spectra for this night in Figures \ref{fig:model_flag_PS_data_SSINS} and \ref{fig:model_flag_PS_data_AOFlagger} on a per-pointing basis. The \textsc{SSINS} flags show much more variation between pointings, which is reflected in the power spectra of Figure \ref{fig:model_flag_PS_data_SSINS}. In contrast, the \textsc{AOFlagger} power spectra show little variation between pointings, and consequently there is little variation between them. In both cases, there is far more power in the EoR window than the expected EoR signal, and intervention is required. 
%3, 2, 6, 4, 5

The window power in the \textsc{AOFlagger}-informed power spectra appear to average down favorably. This may be due to the fact that this is a relatively benign night, and much of the RFI that does exist is probably fainter than the sensitivity of \textsc{AOFlagger}, i.e. the flags are dominated by false positives. In \cite{Offringa2019a}, \textsc{AOFlagger} more clearly reports true positives, and so they see that the excess power from flagging does not average incoherently like thermal noise might be expected to. Instead, since certain frequencies are consistently occupied by certain transmitters, the power does not average down substantially. This is more like what is shown in Figure \ref{fig:model_flag_PS_data_SSINS}, where consistent DTV events are identified by \textsc{SSINS}, resulting in relatively poor average qualities in the EoR window. 

\textsc{AOFlagger} consistently flags the frequency 194.455 MHz about 10 per cent more often than other frequencies, fractionally. This ultimately puts a floor on how far down any averaging can reach, since a narrowband flagging shape will put constant excess power along the line-of-sight. The cause of this consistent flagging is so-far unknown. Since it is within a DTV allocation, it is confounding that the flags would be so narrow. It is close to the DTV guard band, and we do sometimes see \textsc{AOFlagger} flags around the guard band of extremely faint DTV signals observable by \textsc{SSINS}. However, this would suggest constant observation of a DTV signal that is not often identified by \textsc{SSINS}. Furthermore, the guard band flags from \textsc{AOFlagger} usually occupy more than one fine frequency channel, and this is localized to a particular channel. It is possibly locally generated RFI that only affects very few baselines, and thus is observed by \textsc{AOFlagger} but not by \textsc{SSINS}. 

We see that with realistic flagging patterns, even relatively clean data will be too contaminated by excess flagging power for an EoR detection. The low resolution \textsc{SSINS} flags do not reduce the average occupancy by much since the vast majority of the flags are broadband. On the other hand, the \textsc{AOFlagger} flags are sparse, resulting in an order of magnitude decrease in the flag occupancy at low resolution; most of the flags result in partially flagged visibilities. This results in the need for slightly different mitigation strategies for the power resulting from the different sets of flags.

\subsection{Mitigation via Flag Extension}

\begin{figure*}
    \centering
    \begin{subfigure}[]{\includegraphics[width=0.3\linewidth, page=3]{flagging_PS_figs/_2014_GS_broadcast_titles.pdf}
    }
    \end{subfigure}
    \begin{subfigure}[]{\includegraphics[width=0.3\linewidth, page=2]{flagging_PS_figs/_2014_GS_broadcast_titles.pdf}
    }
    \end{subfigure}
    \begin{subfigure}[]{\includegraphics[width=0.3\linewidth, page=6]{flagging_PS_figs/_2014_GS_broadcast_titles.pdf}
    }
    \end{subfigure}
    \begin{subfigure}[]{\includegraphics[width=0.3\linewidth, page=4]{flagging_PS_figs/_2014_GS_broadcast_titles.pdf}
    }
    \end{subfigure}
    \begin{subfigure}[]{\includegraphics[width=0.3\linewidth, page=5]{flagging_PS_figs/_2014_GS_broadcast_titles.pdf}
    }
    \end{subfigure}
    \begin{subfigure}[]{\includegraphics[width=0.3\linewidth, page=1]{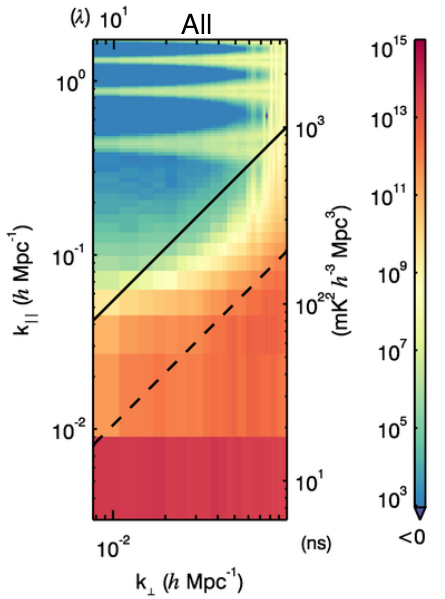}
    }
    \end{subfigure}

    \caption{Simulated per-pointing power spectra using achromatic flags developed from the \textsc{SSINS} flags in \S\ref{sec:data_flags}. The five pointings are shown in chronological order left-to-right, top-to-bottom. See Figure \ref{fig:model_flag_PS_data_SSINS} caption for pointing designations. The sixth power spectrum is the result from integrating the entire night. There is no excess power from these flags, with a relatively small decrease in overall data volume.}
    \label{fig:flag_extend_GS}
\end{figure*}

In Figure \ref{fig:flag_extend_GS}, we show the results of applying a flag extension strategy to the \textsc{SSINS} flags before constructing power spectra. In this case, every integration with an RFI flag anywhere in the band is flagged entirely across the band. We see that this fully eliminates the excess power from flagging at the expense of a modest amount of data. Interestingly, the power could also be fully removed by setting the observing band equal to the occupied band. However, even comparatively broad RFI sources such as DTV are significantly narrower than typical bandwidths used in EoR power spectrum estimates, and changing the observing band to match this width would come at the cost of significant sensitivity.

The sensitivity loss for a flag extending strategy depends on the initial flags. For the night used in \S\ref{sec:data_flags}, the \textsc{SSINS} occupancy is 4.4 per cent, while the \textsc{AOFlagger} occupancy is 2 per cent. Both of these figures ignore routine flagging at the coarse band edges and centers, as well as routine flagging at the beginning and end of observations. When flags are extended over the entire analysis band, the \textsc{SSINS} occupancy approximately doubles, resulting in a modest sensitivity decrease. Since \textsc{AOFlagger} has a false positive rate of order $1 / N_\nu$, where $N_\nu$ is the number of frequencies in the band, extending the flags at the finest resolution results in substantial data loss. After extending the flags at the finest resolution, the effective \textsc{AOFlagger} occupancy is 86 per cent. If instead the high-resolution flags are petrified into the data during downsampling and only the low-resolution flags are extended, then the ensuing flag occupancy is 15 per cent. However, this will leave the high-resolution excess flagging power in the data. Depending on the level of downsampling, one could implement some of the mitigation strategies in \cite{Offringa2019a} on the high-resolution data in order to remedy this. This would allow flags from \textsc{SSINS} and \textsc{AOFlagger} to be combined and still cohere with the flag-extension strategy. Additionally, one could imagine setting a threshold number of frequency channels such that flags are only extended if the number of flagged channels exceeds the threshold. Since false positives are randomly distributed independently throughout the band, this should in principle reduce the unnecessary data loss by some factor that is exponential in the threshold. However, this would demand that some alternative mitigation strategy be employed for the remaining chromatic flags. The exact balance of sensitivity loss and the excess flagging power in a measured power spectrum depends on the true power of the foregrounds in the data, which itself depends on the foreground subtraction strategy, if any. We find that leaving chromatic flags in the data places high demands on foreground subtraction, even for measurements focused in the EoR window.

\begin{figure*}
    \centering
    \includegraphics{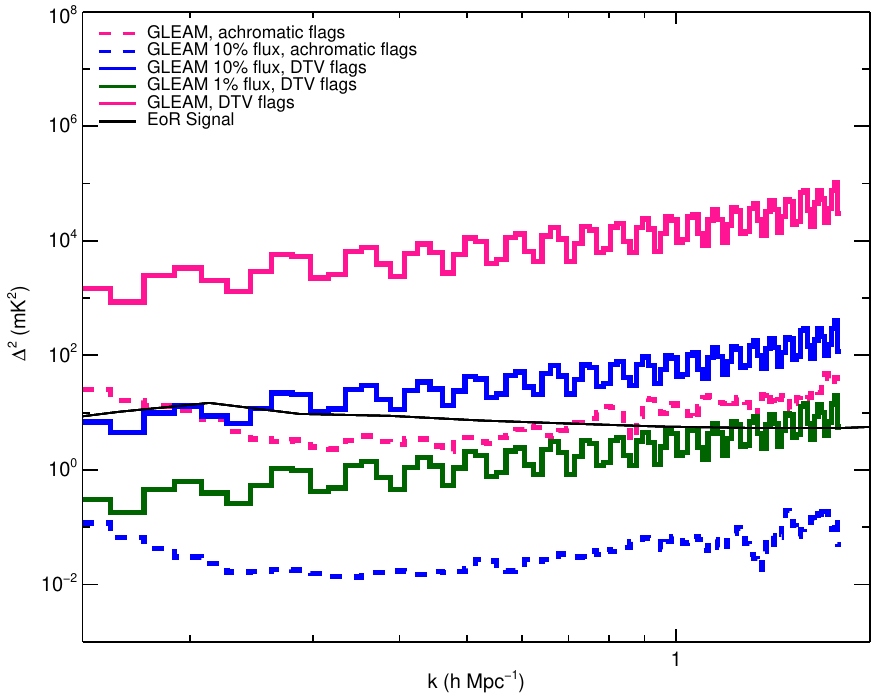}
    \caption{Spherical power spectra computed from simulations using GLEAM sources with various flagging and levels of foreground subtraction, compared to a somewhat optimistic fiducial EoR model (solid black). Coarse band edge flags have been removed for clarity, but would be present in current MWA EoR analyses. Solid lines show simulations with DTV flags, while dashed lines show simulations with achromatic flags. Pink, blue, and green lines show zero, 90 per cent, and 99 per cent foreground removal, respectively. Even with achromatic flags, some subtraction is required for a significant detection of this EoR model. With 90 per cent of the flux removed, chromatic flagging leads to overwhelming power in the modes shown, whereas achromatic reduces the modeled foreground power to manageable levels. A significant detection of the fiducial model for most of these modes would require greater than 99 per cent foreground removal, not including substantial contributions to foreground power from diffuse emission.}
    \label{fig:1d_foreground_sub_flag}
\end{figure*}

\subsection{Forward Model Subtraction}

This excess power is beneath the thermal sensitivity of a small data set, and would require as much data as a season to truly probe in an observational setting. In order to investigate the effect of forward model subtraction, we perform the same two-minute simulations as in \S\ref{sec:flagged_GLEAM}, but with only a subset of the GLEAM catalog. This simulates perfect foreground subtraction up to a certain flux threshold.

To develop the catalog subsets, we rank sources according to their true flux,\footnote{The observing field in question is a relatively quiet field with very few outlying sources. Therefore we expect ranking by true flux and apparent flux to be similar.} and remove the brightest sources from the catalog until 10 per cent and 1 per cent flux levels are achieved. We then simulate two minutes of visibilities for each catalog subset with different flags applied: 1 minute of DTV flags and 1 minute of achromatic flags. After this, we form power spectra and compare. Since the overall contamination levels of the simulations with DTV flags are similar between the two-minute simulations (Figure \ref{fig:flag_PS_sim_GLEAM}(d)) and the full-night simulations (Figure \ref{fig:model_flag_PS_data_SSINS}(f)), we expect the results with a full-night or more to be similar. 

We show spherical power spectra from these simulations in Figure \ref{fig:1d_foreground_sub_flag}. In forming these power spectra, we have implemented a foreground avoidance strategy wherein we only average together bins in the EoR window. What we find is that about 99 per cent of the flux in the GLEAM catalog needs to be removed before the power spectrum resulting from the flagged simulation resembles the unflagged case with no foreground subtraction. For most of the modes shown, this brightness is still too high. We also note that these simulations do not include diffuse structure. This suggests that chromatic flags could only be acceptable if a foreground removal strategy that accurately subtracts greater than 99 per cent of the flux is available. On the other hand, if achromatic flags are used and 90 per cent of GLEAM is subtracted perfectly, the remaining foreground power is sufficiently diminished in the modes shown for typical theoretical EoR signal estimates, with a potential margin for diffuse power. Inaccuracies in the removal strategy can result in EoR signal loss, and so it is imperative to reduce the demands of foreground subtraction as much as possible.

\section{Conclusion}
\label{sec:flag_ps_discussion}

We find that chromatic data flagging produces excess power in the EoR window that in many cases is strong enough to prevent EoR signal detection. This effect arises because flagging causes spectral variations in the sampling function of the array by displacing baseline locations from their unflagged counterparts. The strength of the excess power is determined by both the strength of the foregrounds and the strength of the sampling function disruption. Without mitigation, chromatic flags will prevent a significant EoR detection.

The excess power is produced even in relatively clean data with overall RFI occupancy below 5 per cent. Excess flagging power can occur in two stages. It can be petrified into the data from downsampling visibilities that have flags applied, producing partially flagged data, and it can also result when downsampled visibilities are fully flagged and excluded from the analysis. \cite{Offringa2019a} discusses the case of partially flagged visibilities and provides several effective mitigation strategies. Some of these strategies are either unapplicable or fail when downsampling by only a small factor and when the dominant RFI sources are broad, producing few or no partially flagged samples. Some strategies of \cite{Offringa2019a} do apply to fully flagged samples, but have strict accuracy requirements. We provide an additional mitigation strategy for fully flagged visibilities.

In our investigation, a simple and highly successful mitigation strategy was to extend the flags across the analysis band. This method essentially requires a flagging strategy with a very low false positive rate since the additional flags magnify the false positive rate. If we assume that the RFI brightness distribution is supported down to extremely faint levels, achieving a lower false positive rate by reducing the sensitivity of the flagging algorithm implies a higher false negative rate as well. As shown in \citet{Wilensky2020}, the cost of false negatives can be very high. The optimally sensitive flagging strategy will balance the cost of the false negative rate with the sensitivity loss from extending flags. An accurate evaluation of this cost function requires an understanding of the RFI statistics of the measurement set in terms of brightness and chromaticity.

Since the strength of the excess power depends on the foreground strength, a foreground subtraction strategy can also reduce the excess power. We find that for broad shapes such as DTV, nearly 99 per cent of the total flux on the sky needs to be subtracted perfectly in order to reduce the excess window power to a level comparable to extending the flags. Even at this level, contamination is still too high for constraining most EoR models, meaning substantially deeper foreground subtraction will be required if chromatic flags are used. Inaccurate foreground removal comes with the risk of EoR signal loss, which can lead to artificially low upper limits or false detections of the EoR signal if mischaracterized. Therefore, there is an array of experimental demands that must be balanced including thermal sensitivity, excess power from chromatic flags, the possible necessity for deep foreground subtraction, and the probability of missed RFI.

\section*{Acknowledgements}
We thank the development teams of \textsc{pyuvdata}, \textsc{FHD}, and $\varepsilon$\textsc{ppsilon} which enabled this work. This work was directly supported by NSF grants AST-1643011, AST-1613855, OAC-1835421, and AST-1506024. This project has received funding from the European Research Council (ERC) under the European Union's Horizon 2020 research and innovation programme (grant agreement No.~948764). This scientific work makes use of the Murchison Radio-astronomy Observatory, operated by CSIRO. We acknowledge the Wajarri Yamatji people as the traditional owners of the Observatory site. Support for the operation of the MWA is provided by the Australian Government (NCRIS), under a contract to Curtin University administered by Astronomy Australia Limited. 
 
 \section*{Data Availability}
 The raw data underlying this article are available through the MWA All-Sky Virtual Observatory (ASVO)\footnote{https://asvo.mwatelescope.org}. The \textsc{AOFlagger} flags used for the analysis are also available from ASVO and can be gained simultaneously with the raw data. For instructions on how to find the data on ASVO, please contact the corresponding author. The \textsc{SSINS} flags are available from the corresponding author on reasonable request.

\bibliographystyle{mnras}
\bibliography{main}

%Many modern radio astronomy experiments make use of array interferometers. The measurements, or visibilities, from these instruments sample the fourier dual to the sky brightness map. The fourier domain to the sky is referred to as the $uv$-plane. The sampling function of any given measurement is determined by the relative locations of pairs of receiving elements in the array; pairs are referred to as baselines. This sampling function is the fourier dual of the synthesized point-spread-function (PSF) of the array. In general, arrays with smooth sampling functions with good coverage often have less deleterious PSFs. 

%There are two effects relevant to this work that affect the $uv$-coverage of a given array configuration: rotation synthesis and baseline migration as a function of frequency, e.g. \cite{Morales2012, Morales2019}. The first effect, to enhance coverage is to make use of the Earth's rotation using a technique called rotation synthesis. In short, the sampling function for a given snapshot is determined by the projected locations of the baselines from phase center. Tracking a source across the sky results in baselines tracing out elliptical tracks in the $uv$-plane, thus extending the coverage of the sampling function. 

\end{document}